\documentclass [12pt,a4paper] {article}

\usepackage [latin1] {inputenc} 
\usepackage {url}
\usepackage {color}

\usepackage [colorlinks=true,urlcolor=blue,citecolor=red,linkcolor=blue] {hyperref} 

\begin {document}
\pagestyle {plain}

\title {\bf J. C. Poggendorff: Comment on the Paper by Prof. Kirchhoff (translated and edited by A. K. T. Assis)}
\author {\bf J. C. Poggendorff}
\date {}
\maketitle

{\it Editor's Note: An English translation of J. C. Poggendorff's 1857 paper ``Bemerkung zu dem Aufsatz des Herrn Prof. Kirchhoff'', \cite {poggendorff1857}. This paper is related to Kirchhoff's 1857 paper ``On the motion of electricity in wires'', \cite {kirchhoff1857c} with English translation in \cite {kirchhoff1857a}, and to Weber's 1864 paper ``Electrodynamic measurements, especially on electric oscillations'', \cite {weber1864} with English translation in \cite {weber2020g}. Kirchhoff and Weber were the first to deduce the telegraphy equation by taking into account the self-inductance of the wire. Both of them worked with Weber's electrodynamics. They showed, in particular, that when the resistance of the wire was negligible, the electric wave propagates with light velocity.}
\vskip1cm
Fourth version posted in February 2021 (first version posted in December 2019) at \url {www.ifi.unicamp.br/~assis} and \url {http://arxiv.org/abs/1912.05930}

\pagestyle {myheadings}
\markboth {} {}

\newpage

By J. C. Poggendorff\footnote {\cite {poggendorff1857} related to \cite {kirchhoff1857c} with English translation in \cite {kirchhoff1857a}; and to \cite {weber1864} with English translation in \cite {weber2020g}.}$^,$\footnote {Translated and edited by A. K. T. Assis, \url {www.ifi.unicamp.br/~assis}. I thank L. Hecht for relevant suggestions.}$^,$\footnote {The Notes by H. Weber, the Editor of the fourth volume of Wilhelm Weber's {\it Werke}, are represented by [Note by HW:], while the Notes by A. K. T. Assis are represented by [Note by AKTA:].}$^,$\footnote {[Note by HW:] The above remark by J. C. Poggendorff, to which W. Weber refers in the paper on page 130, \cite [p. 130 of Weber's {\it Werke}] {weber1864}, has been included here because of its historical interest.}

\vskip.5cm

Allow me to add the remark concerning the paper on page 193 of this issue,\footnote {[Note by AKTA:] Johann Christian Poggendorff (1796-1877) is referring to Kirchhoff's paper of 1857 published in page 193 of Volume 100 of the {\it Annalen der Physik und Chemie}, \cite {kirchhoff1857c}, with English translation in \cite {kirchhoff1857a}. Poggendorff edited this Journal from 1824 to 1876, where many of Weber and Kirchhoff's papers were published. The modern {\it Annalen der Physik} is the successor to this Journal.} that when I spoke to Professor W. Weber,\footnote {[Note by AKTA:] W. E. Weber (1804-1891) and C. F. Gauss (1777-1855) invented in 1833 the world's first operational electromagnetic telegraph. It was a 3 km long twin lead connecting Göttingen University, where Weber was Professor of Physics, with the Astronomical Observatory ({\it Sternwarte}), directed by Gauss. It worked based on Faraday's law of induction discovered two years earlier, \cite {faraday1832} with Portuguese translation in \cite {faraday2011-e}.

Twenty four years later, in 1857, Weber and Kirchhoff were the first to derive theoretically the complete telegraphy equation. As pointed out by Poggendorff in this Comment, they worked independently from one another and arrived simultaneously at the same result. Utilizing the modern concepts and usual terminology of circuit theory, we can say that they were the first to take into account not only the capacitance and resistance of the wire, but also its self-inductance.} during his recent stay in Berlin, about Professor Kirchhoff's investigations, Professor Weber showed me a complete treatise on the same subject elaborated by him, which, however, he did not yet intend to submit for printing, because he first wanted to have the results of an experimental investigation that he had undertaken jointly with R. Kohlrausch on this subject.\footnote {Rudolph Kohlrausch (1809-1858) collaborated with Weber on the first measurement of Weber's fundamental constant $c$, which should not be confused with the modern constant $c$ appearing in the textbooks. Weber's constant $c$ is written in the modern International System of Units MKSA as $\sqrt 2/\sqrt {\mu_o\varepsilon_o}$, while the modern constant $c$ is written as $1/\sqrt {\mu_o\varepsilon_o}$. Their experiment was performed in 1854 and they published three papers on this subject in 1855, 1856 and 1857, \cite {weber1855a}, \cite {weberkohlrausch1856} and \cite {kohlrauschweber1857}, with English translations in \cite {weber2019b}, \cite {weberkohlrausch2003} and \cite {kohlrauschweber2019}, respectively. They obtained its value as $4.39450\times 10^8\ m/s$, such that $1/\sqrt {\mu_o\varepsilon_o} = 4.39450\times 10^8/\sqrt 2 = 3.1\times 10^8\ m/s$, that is, essentially the same value as light velocity in vacuum.} Professor Kirchhoff's visit to Berlin a few days later gave him the opportunity to comment on the coincidence of their results --- a coincidence which can be called a pleasant one, as both works, starting from essentially the same basis, have led to identical results. This identity certainly deserves special attention in the case of a subject so little researched as the laws of current formation have been so far.\footnote {[Note by AKTA:] Weber and Kirchhoff deduced the telegraphy equation utilizing Weber's 1846 force law between electrified particles, \cite {weber1846} with English translation in \cite {Weber2007b}. They showed, in particular, that when the conductor had negligible resistance, the velocity of propagation of an electric wave is very nearly equal to the velocity of light in vacuum. This result indicated a direct connection between electromagnetism and optics several years before Maxwell (1831-1879). As Kirchhoff pointed out in the paper which is being discussed here (referring to Weber's constant $c$), \cite [pp. 209-210] {kirchhoff1857c} and \cite [406] {kirchhoff1857a}:

\begin {quote}
The velocity of propagation of an electric wave is here found to be $= c/\sqrt 2$, hence it is independent of the cross section, of the conductivity of the wire, also, finally, of the density of the electricity: its value is 41950 German miles in a second, hence very nearly equal to the velocity of light {\it in vacuo}.
\end {quote}

Kohlrausch, who was collaborating with Weber on some experiments related with the propagation of electromagnetic waves, died in 1858. Weber's work has been delayed in publication and appeared only in 1864. He compared his results with those of Kirchhoff and mentioned Poggendorff's paper on Section 6 of his paper, \cite [Section 6, pp. 130-132 of Weber's {\it Werke}] {weber1864} with English translation in \cite [Section 6] {weber2020g}.}

\vfill\eject

\end {document}